\documentclass[preprint,showpacs,showkeys,preprintnumbers,amsmath,amssymb,superscriptaddress]{revtex4}
\usepackage{graphicx}
\usepackage{dcolumn}
\usepackage{bm}

\begin{document}

\title{Thermodynamical features of Verlinde's approach for a
non-commutative Schwarzschild-anti-deSitter black hole in a broad
range of scales}

\author{S. Hamid Mehdipour}

\email{mehdipour@liau.ac.ir}

\affiliation{Department of Physics, College of Basic Sciences,
Lahijan Branch, Islamic Azad University, P. O. Box 1616, Lahijan,
Iran}

\date{\today}

\begin{abstract}
We try to study the thermodynamical features of a non-commutative
inspired Schwarzschild-anti-deSitter black hole in the context of
entropic gravity model, particularly for the model that is employed
in a broad range of scales, from the short distances to the large
distances. At small length scales, the Newtonian force is failed
because one finds a linear relation between the entropic force and
the distance. In addition, there are some deviations from the
standard Newtonian gravity at large length scales.
\end{abstract}


\maketitle

\section{\label{sec:1}Introduction}
The laws of black hole thermodynamics manifestly show that there is
a profound connection between gravity and thermodynamics \cite{bar}.
Two decades ago, Jacobson exhibited that the Einstein field equation
can be derived from the first law of thermodynamics \cite{jac}.
Recently, Padmanabhan applied the equipartition law of energy and
the holographic principle to make a thermodynamical description of
gravity \cite{pad} (see also \cite{pad2}). Afterwards, Verlinde
suggested a novel idea to interpret the gravity as an entropic force
caused by alterations in the information connected to the position
of massive particles \cite{ver}. He described the notions of inertia
and Newton's law of classical mechanics by joining the gravitational
force with a special kind of force that is emerged from the
thermodynamics on holographic screens. Verlinde's conjecture has
widely been studied in different theoretical frameworks
\cite{ent,ent2,ent3,ent4,ent5,ent6,ent7,ent8,ent9,ent10,ent11,ent12,ent13,ent14,ent15,ent16,ent17,ent18,ent19,ent20,
ent21,ent22,ent23,ent24,ent25,ent26,ent27,ent28}. There are also
some criticisms on the entropic gravity scenario which show a number
of deficiencies of the topic including some open challenges
\cite{comm,comm2,comm3,comm4,comm5,comm6,comm7}.

It is widely expected that underneath the emergent description of
gravitational phenomena there is a deeper layer in which the
fundamental microstructure of a quantum space-time plays an
important role. Therefore, if the origin of gravity is an entropic
force, it is essential within the emergent gravity to understand how
the microscopic scale effects such as non-commutative geometry (NCG)
could be emerged. One of the main references on the universal
character of the quantum gravitational effects in the form of some
non-commutative space-time can be found in Ref.~\cite{nonc}. In a
physically inspired kind of NCG, established upon the coordinate
coherent state technique caused by averaging non-commutative
coordinate fluctuations, one can show that the short distance
behavior of point-like structures is improved
\cite{sma,sma2,sma3,sma4,sma5,sma6,sma7,sma8}. The NCG inspired
model produces a class of solutions of Einstein equations which
contain effects of quantum gravity at small scales. \cite{nic1}.
Lately, some properties of the entropic picture of gravity in the
present of some NCG inspired black holes have been investigated,
e.g., the non-commutative Schwarzschild \cite{meh1},
Reissner-Nordstr\"{o}m \cite{meh2} and Schwarzschild-deSitter (SdS)
\cite{meh3} black holes. Here, it is worth mentioning that
non-commutative point-like sources are also important in the
non-commutative field theory in general, with potentially observable
effects even in atomic physics. These effects are linear in the
non-commutativity parameter \cite{ster}, contrary to the typical
situation in gravity, where such effects are quadratic.

In the NCG inspired model, it has been illustrated that the mean
position of a point-like particle in a non-commutative manifold is
no longer characterized by a Dirac-delta function distribution, but
will be described by a Gaussian distribution of minimal width
$\sqrt{\theta}$, where $\theta$ is the smallest fundamental unit of
an observable area in the non-commutative coordinates, beyond which
coordinate resolution is ambiguous. As a result, the curvature
singularity at the origin of black holes is eliminated and a regular
deSitter (dS) vacuum state is appeared instead. The appearance of a
dS core in the centre of the black hole prohibits its collapse into
a singular region. Indeed, a non-commutative black hole is a
combination of the dS core around the origin with a standard metric
of the black hole far away from the origin. In other words, the
small scale behavior of point-like structures is improved such that
the particle mass $M$, in lieu of being totally localized at a
point, is distributed throughout a region of linear size
$\sqrt{\theta}$ as a smeared-like particle. In fact, due to the
appearance of extreme energies at short distances of a
non-commutative manifold, the effects of manifold quantum
fluctuations become considerable and prohibit any measurements to
find a particle position with an accuracy greater than an inherent
length scale. The smallness of the scale would reveal that
non-commutativity effects can be visible just in extreme energy
phenomena. In a general string theory framework one could presume
that $\sqrt{\theta}$ would naturally not be far from the
four-dimensional Planck scale, $L_{Pl}$. Most of the
phenomenological investigations on non-commutative models have
proposed that we live in a four-dimensional space-time and that the
non-commutative energy scale is about $1-10$ TeV
\cite{hin,hin2,hin3,hin4}, accessible to colliders. However, the
bounds coming from the non-commutative QCD are much stronger, at the
level of $1/\sqrt{\theta}>5\times 10^{14}$ GeV \cite{moci}. Since,
the minimal observable length is not exactly determined through
deduction; therefore the scale is generally presumed as smaller than
the typical scale of the standard model of particle physics, i.e.
only less than $ 10^{-16} cm$. In this paper, we include the
non-commutativity correction in the Schwarzschild-anti-deSitter
(SAdS) metric and find the entropic force for the small and large
scales. Throughout the paper, we will use the definitions $\hbar = c
= k_B = 1$. Also, Greek indices run from 0 to 3.

\section{\label{sec:2}non-commutative SAdS black hole}
The non-commutative SAdS metric is given by \cite{nic2}
\begin{equation}
\label{mat:1}ds^2=-\left(1-\frac{2GM_\theta}{r}-\frac{\Lambda}{3}r^2\right)dt^2+
\left(1-\frac{2GM_\theta}{r}-\frac{\Lambda}{3}r^2\right)^{-1}dr^2+r^2
d\Omega^2,
\end{equation}
where the smeared mass distribution $M_{\theta}$ is found to have
the form
\begin{equation}
\label{mat:2}M_{\theta}=M\left[{\cal{E}}\left(\frac{r}{2\sqrt{\theta}}\right)
-\frac{r}{\sqrt{\pi\theta}}e^{-\frac{r^2}{4\theta}}\right],
\end{equation}
where the Gaussian error function is determined by
${\cal{E}}(x)\equiv 2/\sqrt{\pi}\int_{0}^{x}e^{-t^2}dt$. The
non-commutative SAdS metric is comprised of the non-commutative
black hole solution with a negative cosmological term
$\Lambda=-3/l^2$, where $l$ is the cosmological length associated
with the $\Lambda$. In the limit $r/\sqrt{\theta}\rightarrow\infty$,
we have the standard (commutative) SAdS metric. In the commutative
limit, the Gaussian error function tends to one and the second term
in Eq.~(\ref{mat:2}) will exponentially be reduced to zero and
finally one recovers the standard mass totally localized at a point,
i.e. $M_\theta/ M\rightarrow 1$. However, in the regime that
non-commutative fluctuations are important, i.e.
$r\sim\sqrt{\theta}$, the non-commutative SAdS metric deviates
significantly from the standard one and provides a new physics at
short distances.

\begin{figure}[htp]
\begin{center}
\includegraphics{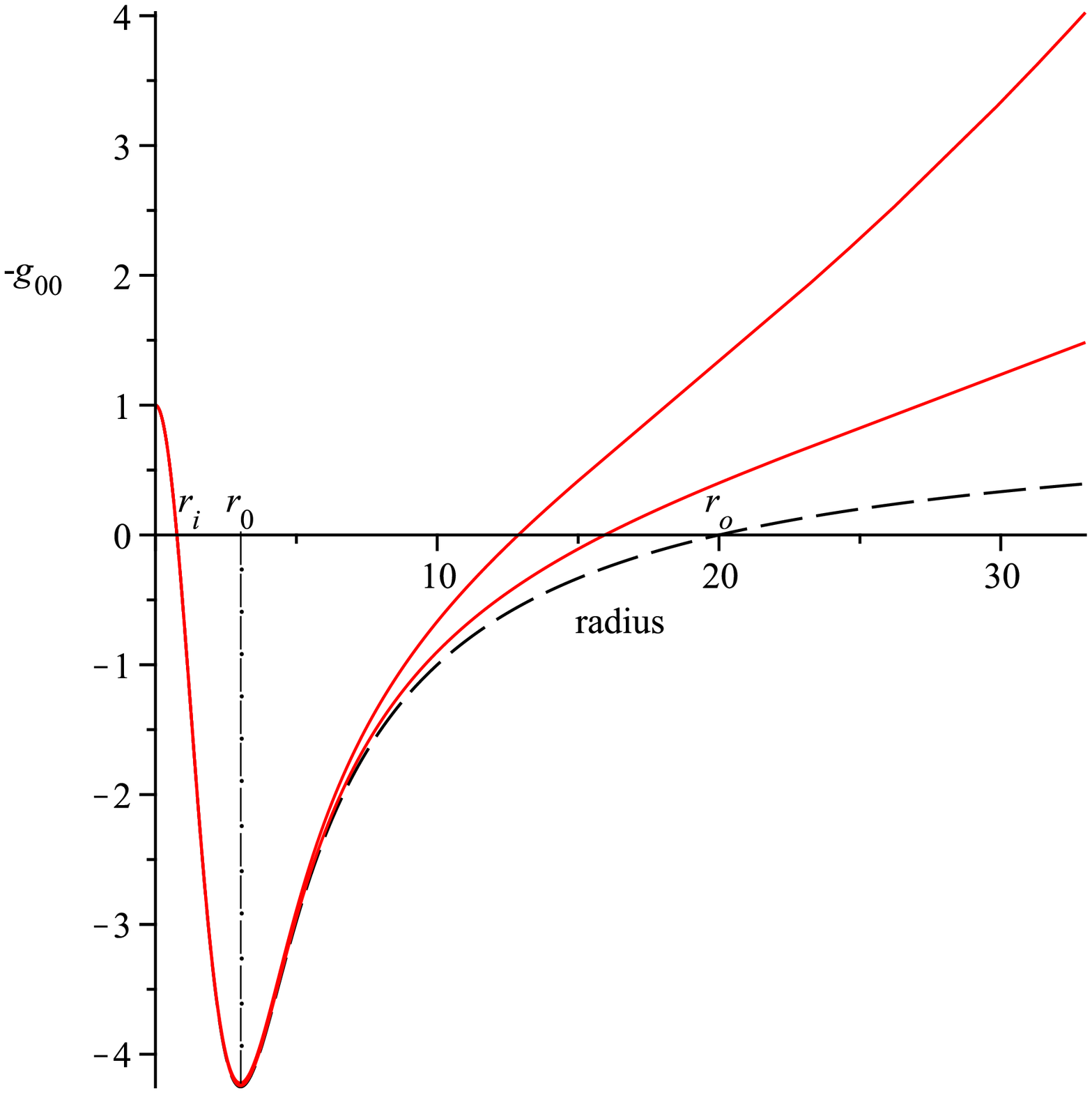}
\end{center}
\vspace{9.9 cm} \caption{\scriptsize {$-\textmd{g}_{00}$ versus the
radius, $r/\sqrt{\theta}$ for $M=10.0\sqrt{\theta}/G$. On the
right-hand side of the figure, from top to bottom, the solid lines
correspond to the non-commutative SAdS black hole for
$\Lambda=-10^{-2}/\theta$, and $\Lambda=-3\times10^{-3}/\theta$,
respectively. The dashed line refers to the non-commutative
Schwarzschild black hole so that it corresponds to $\Lambda=0$. The
radii $r_i$ and $r_0$ are almost fixed for all of the curves, but
the radius $r_o$ is variable associated with $\Lambda$, which is
just shown for the dashed line on the figure.}} \label{fig:1}
\end{figure}
In Fig.~\ref{fig:1}, we present the behavior of $-\textmd{g}_{00}$
versus the radius, $r/\sqrt{\theta}$ for the metric (\ref{mat:1}).
This figure shows two situations. The situation one displays an AdS
background for two values of $\Lambda\theta$ and the situation two
displays an asymptotically flat space. For both situations, there is
no curvature singularity and the metric is regular at the origin.
The possibility of having two distinct horizons for $M>M_0$ is shown
(an inner $r_i$ and an outer black hole horizon $r_o$), where $M_0$
is the minimal mass corresponding to an extremal black hole with one
degenerate horizon in $r_0$. For different values of
$\Lambda\theta$, the minus peak of the curves corresponding to $r_0$
is almost fixed. It is evident from the figure that when the
cosmological constant deviates from the zero, the outer black hole
horizon diminishes however the inner horizon and the minimal
non-zero radius become almost unchanged. In the limit
$\theta\rightarrow 0$, the inner horizon disappears and the outer
horizon is the Schwarzschild value, $r_o = 2M$. However at small
scales or high energies, due to the effect of strong quantum
fluctuations at short distances, quantum gravity corrections
commence to be most significant wherein there is a considerable
deviation from the standard SAdS metric. By expanding
Eq.~(\ref{mat:1}) for $r \gtrsim r_0$, one can find the asymptotic
form of the metric as follows:
\begin{equation}
\label{mat:3}-\textmd{g}_{00}\approx1-\frac{\Lambda_{\textrm{eff}}}{3}r^2,
\end{equation}
with
\begin{equation}
\label{mat:4}\Lambda_{\textrm{eff}}=\Lambda+MG/\sqrt{\pi\theta^3},
\end{equation}
where $r_0$ is a cut-off in the radial direction at small scales and
$\Lambda_{\textrm{eff}}$ is the effective cosmological constant at
short distances. The first expression of Eq.~(\ref{mat:4}) is the
negative background AdS term, while the second expression is the
positive non-commutative fluctuations of the geometry.
\begin{figure}[htp]
\begin{center}
\includegraphics{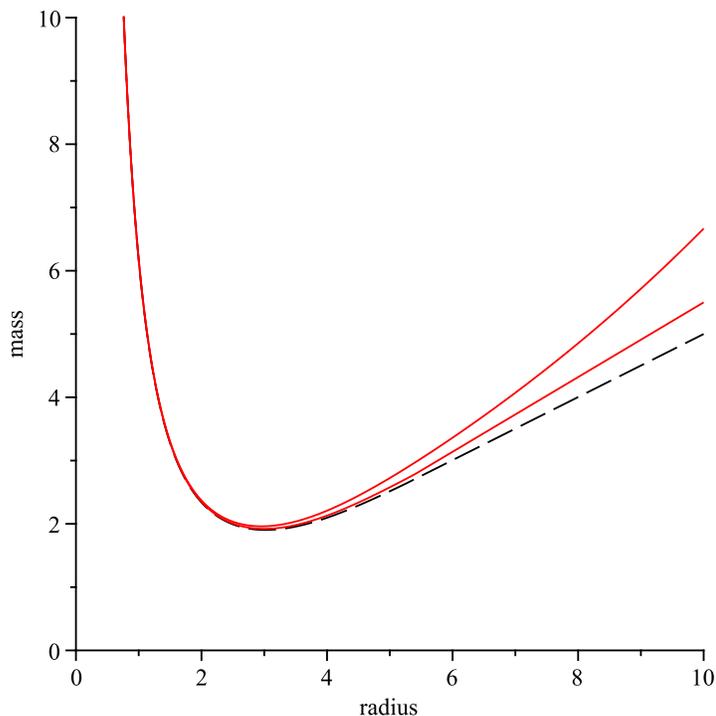}
\end{center}
\vspace{9.5 cm} \caption{\scriptsize { The mass, $M/\sqrt{\theta}$,
versus the radius, $r/\sqrt{\theta}$. On the right-hand side of the
figure, from top to bottom, the solid lines correspond to the
non-commutative SAdS black hole for $\Lambda=-10^{-2}/\theta$, and
$\Lambda=-3\times10^{-3}/\theta$, respectively. The dashed line
refers to the case of $\Lambda=0$.}} \label{fig:1.1}
\end{figure}
It should be emphasized that Eq.~(\ref{mat:3}) is simply obtained
from a Taylor-series-expansion around $r=0$ just to second order in
$r$. We carry out the calculations just under the circumstance that
$r \gtrsim r_0$. The physical description of $r_0$ is the radius of
the smallest holographic surface which can not be probed by a test
particle that is located within some distance from the source. If
one considers the screen radius to be less than the radius of the
smallest holographic surface at the Planckian regime, i.e. $r<r_0$,
then one encounters some unusual dynamical features, leading to
negative entropic force and negative energy \cite{meh1}. As a
consequence, we make the requirement that the screen radius is
bigger than the radius of the smallest holographic surface but is
smaller than the radius corresponding to the maximum extremal
temperature. According to the original work proposed by Nicolini
{\it et al}
\cite{nic3,nic31,nic32,nic33,nic34,nic35,nic36,nic37,nic38,nic39},
for $r\sim\sqrt{\theta}$, the temperature of the black hole grows
during its evaporation until it reaches to a maximum extremal value
and then falls down to a zero temperature black hole remnant
configuration, entirely governed by microscopic fluctuations of the
manifold, encoded in the parameter $\theta$. In other words, instead
of the ordinary divergent treatment for the ultimate phase of the
Hawking evaporation at small radii, there exists a value at which
the temperature vanishes. In addition, for $M<M_0$ there is no
solution for $\textmd{g}_{00}(r_o) = 0$ and no horizon occurs. This
means that, if $r<r_0$ there cannot be a black hole and we cannot
speak of an event horizon and then no temperature can be defined, so
the final zero temperature configuration can be considered a black
hole remnant. To emphasize this point, we consider the internal
energy of the black hole which is nothing but the mass of the black
hole as a function of the event horizon (for more details, see
\cite{nic4}). Following an approach analogous to Ref.~\cite{nic4},
the mass parameter $M$ is a function of the horizon by requiring
$\textmd{g}_{00}(r_o) = 0$. Thus, one can show that there is a
minimum $M_0$ in the following form
\begin{equation}
\label{mat:4.1}M_0\equiv
M(r_0)=\frac{\sqrt{\pi\theta}r_0}{2G}\left[\sqrt{\pi\theta}{\cal{E}}
\left(\frac{r_0}{2\sqrt{\theta}}\right) - r_0
e^{-\frac{r_0^2}{4\theta}}\right]^{-1}\left(1-\frac{\Lambda}{3}r_0^2\right).
\end{equation}
The numerical results of the mass versus the horizon radius are
depicted in Fig.~\ref{fig:1.1}. As can be seen from the figure, the
existence of a minimal non-zero radius ($r=r_0$), corresponding to
the case of an extremal black hole configuration ($M=M_0$), is
clear. For different values of $\Lambda\theta$, the minimal non-zero
radius and the minimal non-zero mass are nearly fixed, i.e.
$r_0\approx3\sqrt{\theta}$ and $M_0\approx1.9\sqrt{\theta}$. As
expected, the non-commutativity discloses a minimal non-zero mass,
namely the black hole remnant, in order to have an event horizon.
So, the black hole in the non-commutative case does not allow to
decay lower than the remnant, and for $M < M_0$ there is no event
horizon.

\section{\label{sec:3}Entropic force at small length scales}
In order to define the temperature, we first need to introduce the
generalized form of the Newtonian potential $\phi$ via the timelike
Killing vector $\xi^\alpha$:
\begin{equation}
\label{mat:6}\phi=\frac{1}{2}\log\left(-\textmd{g}^{\alpha\beta}\xi_\alpha\xi_\beta\right),
\end{equation}
where $\xi_\alpha$ satisfies the Killing equation
\begin{equation}
\label{mat:6.1}
\partial_\alpha\xi_\beta+\partial_\beta\xi_\alpha=2\Gamma^\gamma_{\alpha\beta}\xi_\gamma.
\end{equation}
The redshift factor is denoted by $e^\phi$ and it relates the local
time coordinate to that at a reference point with $\phi = 0$. In
accord with Ref.~\cite{sds}, there is a problem to normalize a
timelike Killing vector in a curved space-time. To study the
entropic force for the non-commutative SAdS metric, which is not an
asymptotically flat space-time, we may consider the normalization of
a timelike Killing vector $\xi_\alpha$, as follows
\begin{equation}
\label{mat:5}\xi_\alpha=\sigma(\partial_0)_\alpha,
\end{equation}
where $\sigma$ is a normalization constant. To find
Eq.~(\ref{mat:5}) we have used the Killing equation and the
condition of static spherical symmetry
$\partial_0\xi_\alpha=\partial_3\xi_\alpha=0$, and also the infinity
condition $\xi_\alpha\xi^\alpha=-1$. It is clear that for the
infinity condition we should have $\sigma = 1$. The gravitational
potential for the non-commutative SAdS metric is then found to be
\begin{equation}
\label{mat:8}\phi=\frac{1}{2}\log\left(-\sigma^2\textmd{g}_{00}\right).
\end{equation}
At small length scales, we use the condition of $r\sim\sqrt{\theta}$
just under the circumstance that
$M>\frac{3\sqrt{\pi\theta^3}}{l^2G}$ (or
$\Lambda_{\textrm{eff}}>0$). For $\Lambda_{\textrm{eff}}>0$ there
are a dS core at the origin and a local gravitational repulsion.
Using the Bousso-Hawking reference point \cite{bou}, one can observe
the temperature on the holographic screens in short distances.
Bousso and Hawking set up a reference point in the radial direction,
wherein the force vanishes. They have indicated that this reference
point can play a role of a point at infinity in an asymptotically
flat space-time. They selected a normalization in which the norm of
the Killing vector is unity at the region where the force vanishes,
the gravitational attraction becomes precisely balanced out by the
cosmological repulsion. Adopting this normalization is associated
with the choosing a special observer who follows geodesics. To find
the correct value for the temperature, one must normalize the
Killing vector in the right way. In the Schwarzschild case ($\Lambda
= 0$) the natural choice is to have $\xi^2 = -1$ at infinity; this
corresponds to $\sigma = 1$ for the standard Schwarzschild metric.
However, in our case there is no infinity, and it would be a mistake
to set $\sigma = 1$. Instead one may choose the radius $r_0$ as a
Bousso-Hawking reference point due to the fact that the temperature
becomes zero at that point. An observer at $r_0$ will need no
acceleration to stay there, just like an observer at infinity in the
Schwarzschild case. One must normalize the Killing vector on this
geodesic orbit. We assume that the region between the inner and
outer black hole horizon is separated by a boundary at the reference
point $r = r_0$. Then the two regions divided by this boundary
cannot have thermal exchange between them because the temperature at
the reference point vanishes wherein a thermally insulating wall is
existed. The notion of thermally insulating wall in our
consideration is similar to that of wholly reflecting wall in the
Gibbons-Hawking's work \cite{gibb}. The two regions separated by the
surface at $r = r_0$ can be thought as independent systems: the
total system becomes the sum of two independent systems, the inner
($r < r_0$) and the outer ($r> r_0$) regions. But, as mentioned
before, we should emphasize that the existence of a dS core in the
centre of the black hole yields a outward push to prevent its
collapse into a singular one. For that reason, it is impossible to
set up a measurement to find more precise particle position than
$r_0$ and for the pattern of the metric for $r < r_0$ no temperature
can be defined.

To characterize the foliation of space, and for recognizing the
holographic surfaces $\Omega$ at screens of the constant redshift,
we should consider the acceleration $a^\alpha$ on the spherical
holographic screen with radius $r$ in a general relativistic form as
\begin{equation}
\label{mat:7}a^\alpha=-\textmd{g}^{\alpha\beta}\nabla_\beta\phi.
\end{equation}
The temperature on the holographic screen seen by an observer
located at the Bousso-Hawking reference point is given by the
Unruh-Verlinde temperature associated with the proper acceleration
of a particle near the screen which is written as \cite{ver}
\begin{equation}
\label{mat:9}T=-\frac{1}{2\pi}e^\phi n^\alpha
a_\alpha=\frac{e^\phi}{2\pi}\sqrt{\textmd{g}^{\alpha\beta}\nabla_\alpha\phi\nabla_\beta\phi},
\end{equation}
where $n^\alpha$ is a unit vector in which it is normal to the
holographic screen and to $\xi_\alpha$. The unit vector is given by
\begin{equation}
\label{mat:10}n^\alpha=\frac{\nabla^\alpha\phi}{\sqrt{\textmd{g}^{\alpha\beta}\nabla_\alpha\phi\nabla_\beta\phi}}.
\end{equation}
The redshift factor $e^\phi$
is identical to unity at the Bousso-Hawking reference point. The
temperature for the non-commutative SAdS metric in short distances
becomes
\begin{equation}
\label{mat:11}T=\frac{\sigma}{4\pi}\left|\frac{d\textmd{g}_{00}}{dr}\right|\approx
\frac{\sigma}{2\pi}\left(\frac{MG}{3\sqrt{\pi\theta^3}}-\frac{1}{l^2}\right)r.
\end{equation}
It is straightforward to show that
\begin{equation}
\label{mat:4.4}\sigma=\left[-\textmd{g}_{00}(r_0)\right]^{-\frac{1}{2}}.
\end{equation}
The energy on the holographic screen according to the equipartition
law of energy can be written as
\begin{equation}
\label{mat:12}E=\frac{1}{4\pi}\int_\Omega e^\phi\nabla\phi dA=2\pi
r^2T,
\end{equation}
where $A$ is the area of the surface. The energy on the
non-commutative SAdS screen is given by
\begin{equation}
\label{mat:13}E\approx\sigma\left(\frac{MG}{3\sqrt{\pi\theta^3}}-\frac{1}{l^2}\right)r^3.
\end{equation}
The modified Newtonian force law as the entropic force is found to
be
\begin{equation}
\label{mat:14}F_\alpha=T\nabla_\alpha S,
\end{equation}
where, the change in entropy for the test mass at a fixed position
near the screen can be written as $\nabla_\alpha S=-2\pi m
n_\alpha$. Note that, the entropy is defined on the freely falling
holographic screen located outside the horizon and the minus sign
for the change in entropy comes from the fact that the entropy
increases when we cross from the outside to the inside. In other
words, we assume that the entire mass distribution is contained
inside the volume enclosed by the holographic screen, and all test
particles are located in the emerged space outside the screen. This
yields the other reason that compels us to consider the holographic
screen located at the distance grater that $r_0$.

We then obtain the entropic force in the presence of the
non-commutative SAdS black hole at small scale,
\begin{equation}
\label{mat:15}F=\sqrt{\textmd{g}^{\alpha\beta}F_\alpha
F_\beta}\approx\frac{\sigma m\Lambda_{\textrm{eff}}}{3}r,
\end{equation}
which is not a Newtonian force. The entropic force vanishes at the
origin and there exists a linear relation between the force and the
distance. The non-Newtonian kind of a entropic force might
potentially be of interest for the domain of validity at small
length scales.

Note that, if we had chosen a different type of the smeared mass
distribution, the overall qualities would be led to completely
similar outcomes to those above. Therefore the fundamental
characteristics of the non-commutativity method are not specifically
sensitive to the Gaussian nature of the smearing. To prove that we
choose a Lorentzian distribution of the smeared mass as follows
\cite{meh1}
\begin{equation}
\label{mat:16}M_{\theta'}=\frac{2M}{\pi}\left[\tan^{-1}\bigg(\frac{r}
{\sqrt{\theta'}}\bigg)-\frac{r\sqrt{\theta'}}{r^2+\theta'}\right].
\end{equation}
The new non-commutativity parameter $\theta'$, is not accurately
similar to $\theta$. In the commutative limit, i.e.
$\theta'\rightarrow0$, one obtains $M_{\theta'}\rightarrow M$. In
the limit $r \gtrsim r'_0$ ($r\sim\sqrt{\theta'}$), by expanding the
following Lorentzian profile of the metric
\begin{equation}
\label{mat:17}ds^2=-\left(1-\frac{2GM_\theta'}{r}-\frac{\Lambda}{3}r^2\right)dt^2+
\left(1-\frac{2GM_\theta'}{r}-\frac{\Lambda}{3}r^2\right)^{-1}dr^2+r^2
d\Omega^2,
\end{equation}
where $r'_0$ is a Lorentzian cut-off in the radial direction at
small scales, we can immediately write the asymptotic form of
$\textmd{g}_{00}$ as
\begin{equation}
\label{mat:18}-\textmd{g}_{00}\approx1-\frac{\Lambda'_{\textrm{eff}}}{3}r^2,
\end{equation}
where $\Lambda'_{\textrm{eff}}=\Lambda+8MG/\pi\sqrt{\theta'^3}$ is
the effective cosmological constant for a Lorentzian distribution in
short distances. For the entropic force, we have finally
\begin{equation}
\label{mat:19}F\approx\frac{\sigma m\Lambda'_{\textrm{eff}}}{3}r,
\end{equation}
which is similar to that found in the Gaussian profile. Hence, most
of the consequences obtained in the Gaussian profile, at least for
asymptotic values of $r$, remain intact if we pick out other
profiles of probability distributions.

\section{\label{sec:4}Entropic force at large length scales}
The time-like Killing vector $\xi_\alpha$ for the non-commutative
SAdS metric which is an asymptotically AdS space-time, can be
achieved by a normalization constant $\sigma$ as
$\xi_\alpha=\sigma(\partial_0)_\alpha$. In an asymptotically flat
space-time, the standard Killing vector normalization, i.e.
$\sigma=1$, is retrieved. Similarly, at large length scales or
$r\gg\sqrt{\theta}$, we use the explicit form of the metric
(\ref{mat:1}) and obtain the Unruh-Verlinde temperature for the
non-commutative SAdS black hole
\begin{equation}
\label{mat:20}T=\frac{\sigma}{2\pi}\left(\frac{GM_{\theta}}{
r^2}-rf(r)\right),
\end{equation}
with
\begin{equation}
\label{mat:21}f(r)=\frac{GM}{2\sqrt{\pi\theta^3}}e^{-\frac{r^2}{4\theta}}-\frac{1}{l^2}.
\end{equation}
Here, it should be noted that the second term of Eq.~(\ref{mat:21})
is extremely small. This means that in the limit
$r\gg\sqrt{\theta}$, the cosmological length scale is extremely
large and an asymptotically flat space-time may approximately be
retrieved at that regime.

The energy on the non-commutative SAdS screen is immediately written
as
\begin{equation}
\label{mat:22}E=\sigma\left(GM_{\theta}-r^3f(r)\right).
\end{equation}
Finally, the entropic force in the presence of the non-commutative
SAdS black hole in large distances becomes
\begin{equation}
\label{mat:23}F=\sigma\left(\frac{GM_{\theta}m}{r^2}-mrf(r)\right).
\end{equation}

\begin{figure}[htp]
\begin{center}
\includegraphics{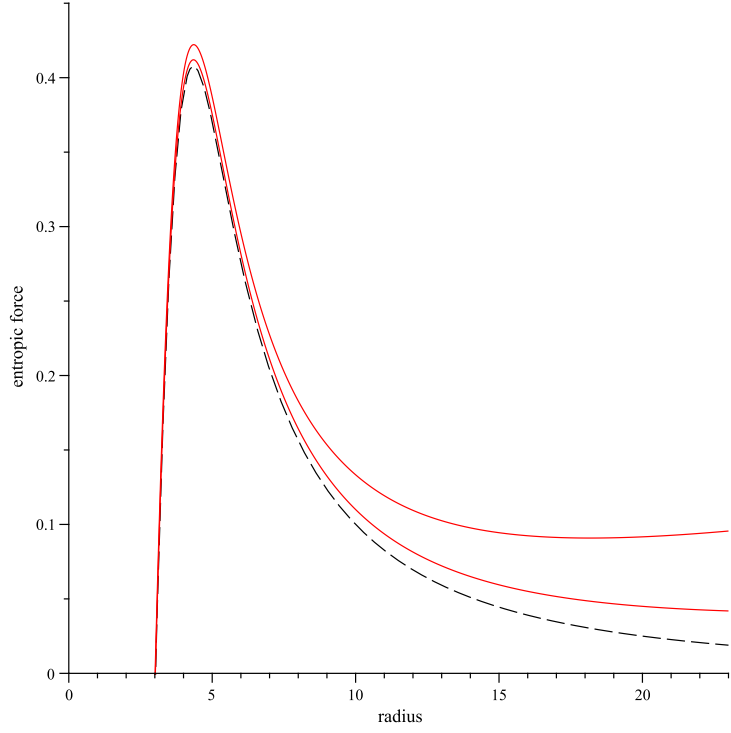}
\end{center}
\vspace{9 cm} \caption{\scriptsize {The entropic force $F$ versus
the radius, $r/\sqrt{\theta}$. We have set $M=10.0\sqrt{\theta}/G$.
On the right-hand side of the figure, from top to bottom, the solid
lines correspond to the non-commutative SAdS black hole for
$\Lambda=-10^{-2}/\theta$, and $\Lambda=-3\times10^{-3}/\theta$,
respectively. The dashed line refers to the non-commutative
Schwarzschild black hole so that it corresponds to $\Lambda=0$.}}
\label{fig:2}
\end{figure}
\begin{figure}[htp]
\begin{center}
\includegraphics{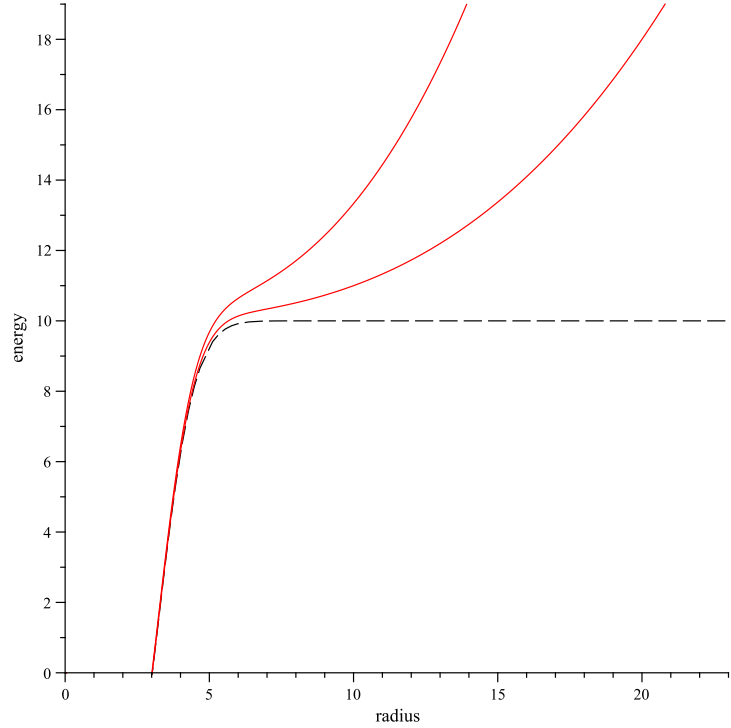}
\end{center}
\vspace{8.5 cm} \caption{\scriptsize {The energy, $E/\sqrt{\theta}$,
versus the radius, $r/\sqrt{\theta}$. We have set
$M=3.0\sqrt{\theta}/G$. On the right-hand side of the figure, from
top to bottom, the solid lines correspond to the non-commutative
SAdS black hole for $\Lambda=-10^{-2}/\theta$, and
$\Lambda=-3\times10^{-3}/\theta$, respectively. The dashed line
refers to the non-commutative Schwarzschild black hole so that it
corresponds to $\Lambda=0$.}} \label{fig:3}
\end{figure}
The numerical computation of the entropic force and the energy as a
function of the radius for two cases, an AdS background and an
asymptotically flat space-time, are depicted in Figs.~\ref{fig:2}
and \ref{fig:3}, respectively. As can be seen from Fig.~\ref{fig:2},
as the cosmological constant deviates from the zero, the entropic
force increases but the peak in the entropic force in the vicinity
of the minimal non-zero radius $r_0$ remains nearly intact.
Similarly, the energy in Fig.~\ref{fig:3} increases with deviating
the cosmological constant from the zero. As we have already
mentioned, the case of $r<r_0$ leads to some out of the standard
dynamical features like negative entropic force, i.e. gravitational
repulsive force, and negative energy; as a result, one should make
the requirement that $E\geq 0$. Accordingly, the appearance of a
lower finite cut-off at the short-scale gravity compels a bound on
any measurements to determine a particle position in a
non-commutative gravity theory.

Notice that one can define the following effective gravitational
constant:
\begin{equation}
\label{mat:24}G_{\textrm{eff}}=G\left[{\cal{E}}\left(\frac{r}{2\sqrt{\theta}}\right)
-\frac{r}{\sqrt{\pi\theta}}e^{-\frac{r^2}{4\theta}}\left(1+\frac{r^2}{2\theta}\right)\right],
\end{equation}
and rewrite the entropic force of the metric (\ref{mat:1}) in terms
of the effective gravitational constant at large length scales as
follows
\begin{equation}
\label{mat:25}F=\sigma\left(\frac{G_{\textrm{eff}}Mm}{r^2}+\frac{m}{l^2}r\right).
\end{equation}
One can easily observe that the NCG inspired model can anticipate an
effective gravitational constant as well. The effective
gravitational constant includes effects of the non-commutativity of
coordinates such that in the limit $r/\sqrt{\theta}\rightarrow
\infty$, we have the standard gravitational constant, i.e.
$G_{\textrm{eff}}\rightarrow G$.

Finally, the last point we should denote here is related to a recent
paper \cite{greg} that analyzed the question of possible quantum
corrections in the entropic scenario of emergent gravity. In our
present work we presume that it is possible to analyze the effects
of the underlying non-commutativity on the entropic gravity via
concept of a smooth commutative holographic screen at small scales.
However, the authors of Ref.~\cite{greg} claim that the holographic
screen in short distances should also be considered as a
non-commutative one. They used a fuzzy sphere as a natural
quasiclassical approximation for the spherical holographic screen to
analyze whether it is possible to observe such corrections to
Newton's law in principle. The main outcome of their analysis is
that it is difficult to draw any conclusive prediction unless there
is a complete control over the dynamics of the microscopic degrees
of freedom leading to the entropic picture.

\section{\label{sec:5}Summary}
In summary, we have studied the thermodynamical aspects of a
non-commutative SAdS black hole in the framework of Verlinde's
conjecture. We have obtained the energy and the entropic force at
small and large scales. The entropic force is linear in $r$ at small
length scales; as a consequence, the Newtonian force is broken down.
Our calculations do not show any severe differences between Gaussian
and Lorentzian profiles. At large length scales, we have found some
deviations from the standard Newtonian gravity. The NCG inspired
model anticipates the presence of an effective gravitational
constant in addition to a lower finite cut-off at the short-scale
gravity.\\

\section*{Acknowledgments}
The author is appreciative of the research council of IAU, Lahijan
Branch for financial assistance. I would like to thank anonymous
referee(s) for useful comments.\\

\end{document}